# An LMI-based Robust Fuzzy Controller for Blood Glucose Regulation in Type 1 Diabetes

Mohammadreza Ganji
*Department of Electrical Engineering*
*Shahid Beheshti University*
*Tehran, Iran*
mganji504@gmail.com

Mahdi Pourgholi
*Department of Electrical Engineering*
*Shahid Beheshti University*
*Tehran, Iran*
m_pourgholi@sbu.ac.ir

*Abstract*— **This paper presents a control algorithm for creating an artificial pancreas for type 1 diabetes, factoring in input saturation for a practical application. By utilizing the parallel distributed compensation and Takagi-Sugeno Fuzzy model, we design an optimal robust fuzzy controller. Stability conditions derived from the Lyapunov method are expressed as linear matrix inequalities, allowing for optimal controller gain selection that minimizes disturbance effects. We employ the minimal Bergman and Tolic models to represent type 1 diabetes glucose-insulin dynamics, converting them into corresponding Takagi-Sugeno fuzzy models using the sector nonlinearity approach. Simulation results demonstrate the proposed controller's effectiveness.**

*Keywords—Linear matrix inequality; Fuzzy control; Robust Control; Vector Lyapunov function; Blood Glucose Regulation; Bergman and Tolic model*

TABLE I. NOMENCLATURE

| Notation | Terms indicating the notation |
|---|---|
| $G$ | Blood sugar concentration above basal value |
| $I$ | Blood insulin level above basal value |
| $X$ | Consistent with the blood insulin level in the remote compartment |
| $d(t)$ | Meal glucose disturbance (disturbance input) |
| $u(t)$ | External insulin infusion rate (control input) |
| $G_b$ | Basal value of blood sugar concentration |
| $I_b$ | Basal value of blood insulin concentration |
| $V_1$ | Insulin distribution volume |
| $N$ | Fractional vanishing rate of the blood insulin |
| $G(t)$ | Glucose concentration |
| $I_p(t)$ | Insulin concentration in the blood |
| $I_i(t)$ | Insulin concentration in the intercellular space |
| $X_i(t)$, $i=1,2,3$ | The time delays – related auxiliary variables |
| $V_p$ | Blood insulin distribution volume |
| $V_i$ | Intercellular space effective volume |
| $t_p$ | Insulin decadence time constant in the blood |
| $t_d$ | Insulin decadence time constant in the intercellular space |
| $G_{in}$ | The exogenous glucose infusion rate |
| $f_1(G)$ | Pancreatic insulin generation regulated by blood sugar determiner |
| $f_2(G)$ | Insulin-independent sugar consumption by nerves and brain |
| $f_3(G)f_4(I_i)$ | Insulin-dependent sugar consumption by fat cells and muscle |
| $f_5(x_3)$ | The function which models the auxiliary variables-related hepatic glucose |
| $i$ | i-th number of the rules |
| $z(t)$ | Premise variable vector |
| $x(t)$ | State vector |
| $u(t)$ | Input vector |
| $v(t)$ | Disturbance input vector |
| $y(t)$ | Output vector |
| $M_{ij}$ | j-th fuzzy membership function related to the i-th premise variable |

## I. INTRODUCTION

Type 1 Diabetes (T1D) is an autoimmune disease that affects many people around the world [1]. The limitations of traditional therapies and the spread of complications due to lack of proper and timely treatment, as well as the difficulties faced by diabetic patients during insulin injection in traditional methods, has increased the need to design and implement an automatic blood glucose system. The advancement of technology and clinical efforts in recent years have led to the rapid growth of significant successes in this field.

In a healthy individual, the oscillation of blood glucose concentration (BGC) within the normal range of 70-180 mg/dl is regulated by the secretion of two hormones, glucagon and insulin, from the α and β cells of the pancreas, respectively. When blood sugar is low, the pancreas releases glucagon to stimulate the liver to transform stored glycogen into glucose and secrete in the blood. On the contrary, to decrease the plasma glucose concentration, insulin produced by $β$ cells allow the insulin-dependent cells to absorb the blood glucose [2,3]. Due to the dysfunction of this system, patients with T1D need exogenous insulin. Without precise management, patients can develop hyperglycemia *(BGC > 180 mg/dl)*, potentially leading to complications such as cerebral stroke, renal malfunction, peripheral vascular disease. On the other hand, low blood glucose (hypoglycemia) can have immediate severe effects such as coma or death [4]. Therefore, individuals with T1D face a life-long optimization problem to minimize hyperglycemia while avoiding hypoglycemia. To characterize the glucose-insulin phenomena in T1D, different models have been proposed [6], including the Bergman minimal model [7], Hovorka model [8], 19-order Sorensen model [9], and Tolic model [10].

Numerous control methods, both linear and nonlinear, have been proposed to manage blood sugar levels [11]. For instance, in [5], a simplified adaptive switching controller using the Bergman minimal model was designed. In [12], a switched LPV controller is proposed. Switching Proportional-Integral-Derivative (PID) control has been explored and tested in [13]. Sliding mode controllers were studied in [14, 15]. Further, linear and nonlinear model predictive controllers (MPC) have been formulated in [16, 17, 18], where MPC can address both physical and performance-based constraints but at the cost of intensive online optimization and the requirement for a reliable model. In [19], Bergman and Tolic model were utilized to design a Takagi-Sugeno (TS) fuzzy model-based controller to overcome these problems. The principal advantage of this approach is the fact that the controller design procedure for nonlinear systems can be simplified to controller design for

Linear systems by applying TS fuzzy approximators [20]. Therefor compared with nonlinear MPC, fuzzy controllers are less complex with a lower amount of numerical computation. However, in [19] the insulin infusion pump constraint is not considered in the controller design procedure, and the suggested LMI for designing PDC controller has tiny feasibility region, and in most cases, it cannot be solved systematically.

In this study, based on the TS fuzzy system and utilizing the parallel distributed compensation concept for controller design with the vector Lyapunov function approach, a robust $H_\infty$ controller that considers pump input saturation is introduced. The LMI-based controller design process is formulated to be systematically solvable with a large feasible region.

## II. METHODS

### A. Mathematical Models of Glucose-Insulin System in T1D

In this section, to describe the glucose-insulin system in T1D, two models will be used, and their TS fuzzy approximations will be derived.

#### 1) Bergman Minimal Model

Several models have been suggested to model the glucose-insulin system in T1D [1]. An ODE-based nonlinear minimal model was suggested by Richard Bergman to characterize dynamics of glucose-insulin system [7]. This model contains a remote compartment in blood and the effective insulin to link glucose concentration to insulin concentration.

The minimal Bergman model is described as follows [21]:
$$\dot{G} = -P_1 G - X(G + G_b) + d(t)$$
$$\dot{I} = -n(I + I_b) + u(t)/V_1 \quad (1)$$
$$\dot{X} = -P_2 X + P_3 I$$

Table 2 gives the values of the parameters.

TABLE II. MINIMAL MODEL PARAMETERS

| parameters | value | unit |
|---|---|---|
| $P_1$ | 0 | $min^{-1}$ |
| $P_2$ | 0.025 | $min^{-1}$ |
| $P_3$ | 0.000013 | $min^{-1}$ |
| $V_1$ | 12 | $L$ |
| $n$ | 0.0926 | $min^{-1}$ |
| $G_b$ | 4.5 | $mmol/L$ |
| $I_b$ | 15 | $mU/L$ |

The controller design procedure requires an optional range for acceptable blood sugar concentration level. The following range is chosen to reduce the risk of hypoglycemia and hyperglycemia [21].
$$a_1 = 60 \leq G + G_b \leq a_2 = 120 \text{ mg/dl} \quad (2)$$

#### 2) Tolic Model

Sturis et al. developed an ODE model to determine whether the action and reaction between glucose and insulin could lead to ultradian oscillations [22]. Afterward, in [9], Tolic et al. simplified the proposed model. In the Tolic model, the effect of insulin on glucose utilization is considered by considering two main negative feedback loops. The proposed model has six states containing three states to model the delay of insulin kinetic and effect on hepatic glucose production. Due to the model's nonlinear behavior, the simplified version was introduced in which can demonstrate the same characteristics as the original model effectively. The Tolic model is in the following form [10]:
$$\dot{I}_p = aI_p + bI_i + cG + d$$
$$\dot{I}_i = eI_p + fI_i$$
$$\dot{G} = gI_i G + hG + kx_3 + lx_3^2 + nx_3^3 + p \quad (3)$$
$$\dot{x}_1 = rI_p - rx_1$$
$$\dot{x}_2 = rx_1 - rx_2$$
$$\dot{x}_3 = rx_2 - rx_3$$

Table 3 gives the values of the parameters.

TABLE III. TOLIC MODEL PARAMETERS

| parameters | value | unit |
|---|---|---|
| a | -0.233 | $min^{-1}$ |
| b | 0.0182 | $min^{-1}$ |
| c | 4.79× $10^{-3}$ | $mUmg^{-1}min^{-1}$ |
| d | -43.9 | $mU^{-1} min^{-1}$ |
| e | 0.0667 | $min^{-1}$ |
| f | -0.0282 | $min^{-1}$ |
| g | -9.44×$10^{-5}$ | $mU^{-1} min^{-1}$ |
| h | 2.64×$10^{-3}$ | $min^{-1}$ |
| k | 17.5 | $mgmU^{-1}min^{-1}$ |
| l | -0.315 | $mgmU^{-2}min^{-1}$ |
| n | 1.48× $10^{-3}$ | $mgmU^{-3}min^{-1}$ |
| p | 80.5 | $mgmin^{-1}$ |
| r | 0.0833 | $min^{-1}$ |

### B. TS Fuzzy Model

Local linear representations of a nonlinear system in each fuzzy implication can be expressed by fuzzy IF-THEN rules [23]:

if $z_1(t)$ is $M_{i1}$ and ... and $z_p(t)$ is $M_{ip}$, then
$$\begin{cases} \dot{x}(t) = A_i x(t) + B_i u(t) + E_i v(t) \\ \quad\quad y(t) = C_i x(t) \end{cases} \quad (4)$$

The overall TS fuzzy system by employing the singleton fuzzifier and the center average defuzzifier is as follows:
$$\dot{x}(t) = \sum_{i=1}^{r} h_i(z(t))\{A_i x(t) + B_i u(t) + E_i v(t)\}$$
$$y(t) = \sum_{i=1}^{r} h_i(z(t))\{C_i x(t)\} \quad (5)$$
$$\sum_{i=1}^{r} h_i z(t) = 1$$

#### 1) Fuzzy Approximation of the Bergman Model

To obtain the TS fuzzy approximate of Bergman system (1), considering $u^* = -nI_b + u/V_1$, and the upper and lower bounds of $(G+G_b)$ are as stated in (2). Now by employing the SNA to the nonlinear terms of the model, the two-rule TS fuzzy model of minimal Bergman model is:
$$\dot{x}(t) = \sum_{i=1}^{2} h_i(z(t))\{A_i x(t) + B_i u(t) + E_i v(t)\} \quad (6)$$
where $x = [G\ I\ X]^T$: the state vector and

$$A_1 = \begin{bmatrix} P_1 & 0 & -a_1 \\ 0 & n & 0 \\ 0 & P_3 & -P_2 \end{bmatrix}, B_1 = \begin{bmatrix} 0 \\ 1 \\ 0 \end{bmatrix}, E_1 = \begin{bmatrix} 1 \\ 0 \\ 0 \end{bmatrix} \quad (7)$$

$$A_2 = \begin{bmatrix} P_1 & 0 & -a_2 \\ 0 & n & 0 \\ 0 & P_3 & -P_2 \end{bmatrix}, B_2 = \begin{bmatrix} 0 \\ 1 \\ 0 \end{bmatrix}, E_2 = \begin{bmatrix} 1 \\ 0 \\ 0 \end{bmatrix} \quad (8)$$

$$h_1 = \frac{120 - (G + G_b)}{120 - 60}; h_2 = 1 - h_1 \quad (9)$$

#### 2) Fuzzy Approximation of Tolic Model

By adding insulin infusion and external meal to the system (3) and introducing a new variable $G' = G + G_{op}$ ($G_{op} = 80$), which leads to converging system's output to zero [19]. The nonlinear model is described as follows:

$$\begin{bmatrix} \dot{I}_p \\ \dot{I}_i \\ \dot{G'} \\ \dot{x}_1 \\ \dot{x}_2 \\ \dot{x}_3 \end{bmatrix} = \begin{bmatrix} a & b & c & 0 & 0 & 0 \\ e & f & 0 & 0 & 0 & 0 \\ 0 & gG' + gG_{op} & h & 0 & 0 & k + lx_3 + nx_3^2 \\ r & 0 & 0 & -r & 0 & 0 \\ 0 & 0 & 0 & r & -r & 0 \\ 0 & 0 & 0 & 0 & r & -r \end{bmatrix}$$  (10)

$$\begin{bmatrix} I_p \\ I_i \\ G' \\ x_1 \\ x_2 \\ x_3 \end{bmatrix} + \begin{bmatrix} u^* \\ 0 \\ v^* \\ 0 \\ 0 \\ 0 \end{bmatrix}$$

where $u^* = cG_{op} + d - 1/b_u\, u$, and $v^* = hG_{op} + p + d(t)$.

By assuming the range $-30 \leq G' \leq 30$ mg/dL and $-10 \leq x_3 \leq 10$ mU to apply SNA, the derived four-rule Takagi-Sugeno fuzzy model of Tolic model is [19]:

$\dot{x} = \sum_{i=1}^{4} h_i \{A_i x + B_i u^* + E_i v^*\}$  (11)

where

$$A_i = \begin{bmatrix} a & b & c & 0 & 0 & 0 \\ e & f & 0 & 0 & 0 & 0 \\ 0 & M_{11} + gG_{op} & h & 0 & 0 & k + M_{2(i)} \\ r & 0 & 0 & -r & 0 & 0 \\ 0 & 0 & 0 & r & -r & 0 \\ 0 & 0 & 0 & 0 & r & -r \end{bmatrix}, B_i = \begin{bmatrix} 1 \\ 0 \\ 0 \\ 0 \\ 0 \\ 0 \end{bmatrix}, E_i = \begin{bmatrix} 1 \\ 0 \\ 0 \\ 0 \\ 0 \\ 0 \end{bmatrix}, i=\{1,2\};$$

$$A_j = \begin{bmatrix} a & b & c & 0 & 0 & 0 \\ e & f & 0 & 0 & 0 & 0 \\ 0 & M_{12} + gG_{op} & h & 0 & 0 & k + M_{2(j-2)} \\ r & 0 & 0 & -r & 0 & 0 \\ 0 & 0 & 0 & r & -r & 0 \\ 0 & 0 & 0 & 0 & r & -r \end{bmatrix}, B_i = \begin{bmatrix} 1 \\ 0 \\ 0 \\ 0 \\ 0 \\ 0 \end{bmatrix},$$

$$E_i = \begin{bmatrix} 1 \\ 0 \\ 0 \\ 0 \\ 0 \\ 0 \end{bmatrix}, j = \{3,4\};$$

$h_1 = \left(\frac{M_{12} - gG}{M_{12} - M_{11}}\right)\left(\frac{M_{22} - lx_3 - nx_3^2}{M_{22} - M_{21}}\right)$, $h_2 = \left(\frac{M_{12} - gG}{M_{12} - M_{11}}\right)\left(\frac{-M_{21} + lx_3 + nx_3^2}{M_{22} - M_{21}}\right)$,

$h_3 = \left(\frac{-M_{12} + gG}{M_{12} - M_{11}}\right)\left(\frac{M_{22} - lx_3 - nx_3^2}{M_{22} - M_{21}}\right)$, $h_4 = \left(\frac{-M_{12} + gG}{M_{12} - M_{11}}\right)\left(\frac{-M_{21} + lx_3 + nx_3^2}{M_{22} - M_{21}}\right)$

where $M_{11} = -0.0057$, $M_{12} = 0.0057$, $M_{21} = -3.01$ and $M_{22} = 3.29$.

### C. Controller Design Procedure

*1) Lemma1. Congruence Transformation*

Consider symmetric matrix $A$ and non-singular quadratic matrix $T$, $T'AT$ is called the congruence transformation of matrix $A$ and $A<0$, if and only if $T'AT<0$  (12)

The fuzzy parallel distributed compensation simplifies the controller design for a nonlinear system to the controller design for each linear subsystem of the TS fuzzy model of the nonlinear model. In this approach, the TS fuzzy model and each control rule have an equal premise part, and a state feedback controller is used in each consequent part. The $i$-th implication of the fuzzy controller is determined as [23]:

if $z_1(t)$ is $M_{i1}$ and ... and $z_p(t)$ is $M_{ip}$, then
$u_i(t) = K_i x(t)$  (13)

To derive to overall PDC controller (14), the singleton fuzzifier and the center average defuzzifier are employed [23].

$u = \sum_{i=1}^{r} h_i K_i x$  (14)

The closed-loop system containing the PDC controller is as follows [24]:

$\dot{x}(t) = \sum_{i=1}^{r} \sum_{j=1}^{r} h_i h_j \{(A_i + B_i K_i)x(t) + E_i v(t)\}$  (15)

*2) Lemma2. Bounded real lemma*

Consider the LTI system (9) and the control law (16); the aim is to find a state-feedback gain $K_i$.

$u_i(t) = K_i x(t)$  (16)

For a given scalar $\gamma_i > 0$, $\|T_{yv}\|_\infty < \gamma_i$ is enforced if there is a symmetric positive definite matrix $P_i$ ensures that the LMI (37) holds:

$$\begin{bmatrix} A'_{i(cl)} P_i + P_i A_{i(cl)} & B_{i(cl)} & C'_{i(cl)} \\ B'_{i(cl)} & -\gamma_i I & D'_{(cl)} \\ C_{(cl)} & D_{(cl)} & -\gamma_i I \end{bmatrix} < 0, P_i > 0$$  (17)

where $A_{i(cl)} = (A_i + B_i K_i)$, By substituting in (17), we have

$$\begin{bmatrix} (A_i + B_i K_i)' P_i + P_i (A_i + B_i K_i) & E_i P_i & C'_i \\ E'_i P_i & -\gamma_i I & D'_i \\ C_i & D_i & -\gamma_i I \end{bmatrix} < 0,$$  (18)

$P_i > 0$

Due to the multiplication of $K_i$ (as controller gain) and $P_i$ (as decision variable) the inequality (18) is not an LMI, therefore by using Lemma 1, we have:

$$\begin{bmatrix} P_i^{-1} & 0 & 0 \\ 0 & 1 & 0 \\ 0 & 0 & 1 \end{bmatrix} \times$$

$$\begin{bmatrix} (A_i + B_i K_i)' P_i + P_i (A_i + B_i K_i) & E_i P_i & C'_i \\ E'_i P_i & -\gamma_i I & D'_i \\ C_i & D_i & -\gamma_i I \end{bmatrix} \times$$

$$\begin{bmatrix} P_i^{-1} & 0 & 0 \\ 0 & 1 & 0 \\ 0 & 0 & 1 \end{bmatrix} < 0, P_i^{-1} P_i P_i^{-1} > 0$$  (19)

By $X_i = P_i^{-1}, M_i = K_i P_i^{-1}$, the inequality (17) becomes the LMI (20)

$$\begin{bmatrix} (A_i X_i + B_i M_i)' + (A_i X_i + B_i M_i) & * & * \\ E'_i & -\gamma_i I & * \\ C_i X & D_i & -\gamma_i I \end{bmatrix} < 0,$$  (20)

$X_i > 0$

and the controller gain is $K_i = M_i X_i^{-1}$.

*3) Theorem 1. Constraint on the control signal*

Suppose that the initial condition $x(0)$ is defined. The constraint $\|u_i\|_2 \leq \mu_i$ is satisfied for $t \geq 0$ if the LMIs (21, 22) hold [23]

$$\begin{bmatrix} 1 & x(0)' \\ x(0) & X_i^{-1} \end{bmatrix} \geq 0,$$  (21)

$$\begin{bmatrix} X_i & M'_i \\ M_i & \mu_i^2 I \end{bmatrix} \geq 0.$$  (22)

where $X_i = P_i^{-1}$, and $M_i = K_i P_i^{-1}$.

*a) Proof*

Consider $V(x(t)) = x^T(t) P x(t)$ as a Lyapunov function and

$x^T(t) P x(t) \leq 1$.  (23)

Suppose $X = P^{-1}$, then,

$1 - x^T(0) X^{-1} x(0) \geq 0,$  (24)

Which can be transformed into (21) using the Schur complement lemma.

From $\|u(t)\|_2 \leq \mu$,

$u^T(t) u(t) = \sum_{i=1}^{r} \sum_{j=1}^{r} h_i h_j x^T(t) F_i^T F_j x(t) \leq \mu^2$  (25)

thus,

$\frac{1}{\mu^2}\sum_{i=1}^{r}\sum_{j=1}^{r}h_i h_j x^T(t)F_i^T F_j x(t) \leq 1$ (26)

Because (27)
$x^T(t)X^{-1}x(t) < x^T(0)X^{-1}x(0) \leq 1$ for $t>0$,
If $\frac{1}{\mu^2}\sum_{i=1}^{r}\sum_{j=1}^{r}h_i h_j x^T(t)F_i^T F_j x(t) \leq x^T(t)X^{-1}x(t)$, (28)

Then (26) holds. Hence,
$\sum_{i=1}^{r}\sum_{j=1}^{r}h_i h_j x^T(t)\left[\frac{1}{\mu^2}F_i^T F_j - X^{-1}\right]x(t) \leq 0$ (29)

and

$\frac{1}{2}\sum_{i=1}^{r}\sum_{j=1}^{r}h_i h_j x^T(t)\left[\frac{1}{\mu^2}F_i^T F_j + \frac{1}{\mu^2}F_j^T F_i - 2X^{-1}\right]x(t)$
$= \frac{1}{2}\sum_{i=1}^{r}\sum_{j=1}^{r}h_i h_j x^T(t) \times \left[\frac{1}{\mu^2}(F_i^T F_i + F_j^T F_j) - (F_i^T - F_j^T)(F_i - F_j) - 2X^{-1}\right]x(t) \leq$
$\frac{1}{2}\sum_{i=1}^{r}\sum_{j=1}^{r}h_i h_j x^T(t)\left[\frac{1}{\mu^2}(F_i^T F_i + F_j^T F_j) - 2X^{-1}\right]x(t)$
$=\sum_{j=1}^{r}h_i x^T(t)\left[\frac{1}{\mu^2}F_i^T F_i - X^{-1}\right]x(t)$ (30)

If $\frac{1}{\mu^2}F_i^T F_i - X^{-1} \leq 0$, (29) holds. By considering $M_i = F_i X$, we have
$\frac{1}{\mu^2}M_i^T M_i - X \leq 0$ (31)

Deploying Schur complement lemma, LMI (22) can be obtained.

### D. Stability Analysis

By using the concept of vector Lyapunov function [25], the following Lyapunov function is selected:
$V(t)=\sum_{j=1}^{r}h_j v_j(t)=\sum_{j=1}^{r}h_j x_j^T p_j x_j$ (32)

where $p_j>0$. The derivative of the Lyapunov function is:

$\dot{V}=\sum_{j=1}^{r}h_j \dot{v}_j=\sum_{j=1}^{r}h_j(\dot{X}_j^T P_j X_j + X_j^T P_j \dot{X}_j)$ (33)
$=\sum_{j=1}^{r}h_j\left[(A_j X_j + B_j K_j X_j)^T P_j X_j + X_j^T P_j(A_j X_j + K_j X_j)\right]$ (34)
$=\sum_{j=1}^{r}h_j\left[(X_j^T A_j^T + x_j^T K_j^T B_j^T)P_j X_j + X_j^T P_j(A_j X_j + B_j K_j X_j)\right]$ (35)
$=\sum_{j=1}^{r}h_j(X_j^T A_j^T P_j X_j + X_j^T K_j^T B_j^T P_j X_j + X_j^T P_j A_j X_j + X_j^T P_j B_j K_j X_j)$ (36)
$=\sum_{j=1}^{r}h_j\left[X_j^T(A_j^T P_j + K_j^T B_j^T P_j + P_j A_j + P_j B_j K_j)X_j\right]$ (37)

Considering (18), $(A_i + B_i K_i)'P_i + P_i(A_i + B_i K_i)<0$, thus
$A_j^T P_j + K_j^T B_j^T P_j + P_j A_j + P_j B_j K_j <0$, (38)
and $\sum_{j=1}^{r}h_j=1$, thus
$\sum_{j=1}^{r}h_j\left[X_j^T(A_j^T P_j + K_j^T B_j^T P_j + P_j A_j + P_j B_j K_j)X_j\right]<0$ (39)

So $\dot{V}(t)<0$, and from the Lyapunov theorem, the system (15) is asymptotically stable.

To minimize the effects of disturbance on the output signal, considering $H_\infty$ norm of $y$ with respect to $v$ as
$\|T_{yv}\|_\infty = \frac{\|y\|_2}{\|v\|_2} < \gamma_i$ (40)

Therefore, sufficient condition for $H_\infty$ performance is that,
$\dot{V} + y^T y - \gamma_i^2 < 0$ (41)

## III. SIMULATION RESULTS

In this section, by applying theorem 1 and lemma 2 simultaneously, the local feedback gains considering different ranges of input saturation will be obtained, then the control signal will be exerted to the nonlinear Bergman and Tolic models. The *Yalmip* toolbox and *Mosek* solver are utilized to solve LMI-based controller design conditions.

### A. Minimal Bergman model

Based on the local LTI subsystems of the Bergman model, the state feedback controller gains considering three different range for the upper bound of the control signal are derived as:

If we consider $0 \leq u \leq 6$, the following Lyapunov matrices and feedback gains are (controller 1):
$\mu=0.095$;
$X_1 = 10^3 \times \begin{bmatrix} 0.009 & -0.106 & 0.0 \\ -0.106 & 1.503 & -0.0002 \\ 0.0 & -0.0002 & 0.0 \end{bmatrix}$, $K_1 = \begin{bmatrix} 0.16 \\ -0.03 \\ -258.18 \end{bmatrix}$, $\gamma_1 = 50.5$;

$X_2 = \begin{bmatrix} 7.443 & -44.47 & 0.007 \\ -0.106 & 1.503 & -0.055 \\ 0.007 & -0.055 & 0.00 \end{bmatrix}$, $K_2 = \begin{bmatrix} 0.179 \\ -0.051 \\ -477.65 \end{bmatrix}$, $\gamma_2 = 36.3$;

If we consider $0 \leq u \leq 10$, the following Lyapunov matrices and feedback gains are (controller 2):
$\mu=0.25$;
$X_1 = 10^3 \times \begin{bmatrix} 0.006 & -0.086 & 0.0 \\ -0.086 & 1.487 & -0.0002 \\ 0.0 & -0.0002 & 0.00 \end{bmatrix}$, $K_1 = \begin{bmatrix} 0.47 \\ -0.06 \\ -597.9 \end{bmatrix}$, $\gamma_1 = 32.3$;

$X_2 = \begin{bmatrix} 5.61 & -39.03 & 0.006 \\ -39.03 & 412.24 & -0.056 \\ 0.006 & -0.056 & 0.0 \end{bmatrix}$, $K_2 = \begin{bmatrix} 0.5 \\ -0.1 \\ -1000.2 \end{bmatrix}$, $\gamma_2 = 24.6$.

If we consider $0 \leq u \leq 25$, the following Lyapunov matrices and feedback gains are (controller 3):
$\mu=1.2$;
$X_1 = 10^3 \times \begin{bmatrix} 0.004 & -0.074 & 0.0 \\ -0.074 & 2.190 & -0.0003 \\ 0.0 & -0.0003 & 0.00 \end{bmatrix}$, $K_1 = \begin{bmatrix} 2.4 \\ 2 \\ -1983 \end{bmatrix}$, $\gamma_1 = 17$;

$X_2 = \begin{bmatrix} 3.99 & -39.30 & 0.005 \\ -39.30 & 785.85 & -0.08 \\ 0.005 & -0.08 & 0.0 \end{bmatrix}$; $K_2 = \begin{bmatrix} 2.4 \\ 2 \\ -3291.5 \end{bmatrix}$, $\gamma_2 = 13.9$.

### B. Tolic model

Based on the local LTI subsystems of the Tolic model, the state feedback controller gains considering 3 different range for the upper bound of the control signal are derived as:

If we consider $0 \leq u \leq 6$, the following Lyapunov matrices and feedback gains are (controller 1):
$\mu=0.004$;
$X_1 = \begin{bmatrix} 52.6 & -19.3 & 31.5 & -24.8 & 8.7 & -1.9 \\ -19.3 & 7.5 & -10.8 & 9 & -3.1 & 0.6 \\ 31.5 & -10.8 & 51.6 & -15.7 & 6.3 & -1.8 \\ -24.8 & 9 & -15.7 & 11.7 & -4.1 & 0.9 \\ 8.7 & -3.1 & 6.3 & -4.1 & 1.5 & -0.3 \\ -1.9 & 0.6 & -1.8 & 0.9 & -0.3 & 0.08 \end{bmatrix}$,

$K_1 = \begin{bmatrix} -0.1 \\ -0.02 \\ -0.01 \\ -0.3 \\ -0.6 \\ -1.1 \end{bmatrix}$, $\gamma_1 = 57.5$;

$X_2 = \begin{bmatrix} 25.9 & -9.8 & 20.7 & -12 & 4.1 & -0.8 \\ -9.8 & 4 & -7.2 & 4.5 & -1.5 & 0.3 \\ 20.7 & -7.2 & 45.9 & -10.3 & 4.1 & -1.1 \\ -12 & 4.5 & -10.3 & 5.6 & -1.9 & 0.4 \\ 4.1 & -1.5 & 4.1 & -1.9 & 0.7 & -0.1 \\ -0.8 & 0.3 & -1.1 & 0.4 & -0.1 & 0.03 \end{bmatrix}$; $K_2 = \begin{bmatrix} -0.1 \\ -0.02 \\ -0.01 \\ -0.4 \\ -0.8 \\ -1.4 \end{bmatrix}$,
$\gamma_2 = 50.3$;

$$X_3 = \begin{bmatrix} 52.4 & -19.2 & 31.4 & -24.7 & 8.7 & -1.8 \\ -19.2 & 7.5 & -10.8 & 8.9 & -3 & 0.6 \\ 31.4 & -10.8 & 51.5 & -15.7 & 6.3 & -1.8 \\ -24.7 & 8.9 & -15.7 & 11.7 & -4.1 & 0.9 \\ 8.7 & -3 & 6.3 & -4.1 & 1.5 & -0.3 \\ -1.8 & 0.6 & -1.8 & 0.9 & -0.3 & 0.08 \end{bmatrix};$$

$$K_3 = \begin{bmatrix} -0.1 \\ -0.02 \\ -0.01 \\ -0.3 \\ -0.6 \\ -1.1 \end{bmatrix}, \gamma_3 = 57.5;$$

$$X_4 = \begin{bmatrix} 24.6 & -9.3 & 19.7 & -11.4 & 3.9 & -0.8 \\ -9.3 & 3.8 & -6.9 & 4.2 & -1.4 & 0.3 \\ 19.7 & -6.9 & 43.8 & -9.8 & 3.9 & -1.1 \\ -11.4 & 4.2 & -9.8 & 5.3 & -1.8 & 0.4 \\ 3.9 & -1.4 & 3.9 & -1.8 & 0.6 & -0.1 \\ -0.84 & 0.3 & -1.1 & 0.4 & -0.1 & 0.03 \end{bmatrix}, K_4 = \begin{bmatrix} -0.1 \\ -0.02 \\ -0.01 \\ -0.4 \\ -0.8 \\ -1.4 \end{bmatrix},$$

$\gamma_4 = 48.1.$

If we consider $0 \leq u \leq 12$, the following Lyapunov matrices and feedback gains are (controller 2):

$\mu = 0.08;$

$$X_1 = \begin{bmatrix} 27.4 & -11.7 & 10.4 & -12 & 3.7 & -0.7 \\ -11.7 & 6.2 & -3.9 & 5 & -1.5 & 0.2 \\ 10.4 & -3.9 & 11.9 & -5 & 1.8 & -0.4 \\ -12 & 5 & -5 & 5.3 & -1.7 & 0.3 \\ 3.7 & -1.5 & 1.8 & -1.7 & 0.5 & -0.1 \\ -0.7 & 0.2 & -0.4 & 0.3 & -0.1 & 0.02 \end{bmatrix},$$

$$K_1 = \begin{bmatrix} -0.4 \\ -0.006 \\ -0.2 \\ -2.8 \\ -8.9 \\ -21.5 \end{bmatrix}, \gamma_1 = 11.4;$$

$$X_2 = \begin{bmatrix} 16.9 & -7.7 & 7.9 & -7.2 & 2.1 & -0.3 \\ -7.7 & 4.6 & -3.1 & 3.2 & -0.9 & 0.1 \\ 7.9 & -3.1 & 11.4 & -3.7 & 1.3 & -0.3 \\ -7.2 & 3.2 & -3.7 & 3.1 & -0.9 & 0.1 \\ 2.1 & -0.9 & 1.3 & -0.9 & 0.3 & -0.06 \\ -0.3 & 0.1 & -0.3 & 0.1 & -0.06 & 0.01 \end{bmatrix},$$

$$K_2 = \begin{bmatrix} -0.5 \\ -0.002 \\ -0.2 \\ -3.3 \\ -11.5 \\ -29.5 \end{bmatrix}, \gamma_2 = 10.5;$$

$$X_3 = \begin{bmatrix} 27.32 & -11.7 & 10.4 & -12 & 3.7 & -0.7 \\ -11.7 & 6.2 & -3.9 & 5 & -1.5 & 0.2 \\ 10.4 & -3.9 & 11.9 & -4.9 & 1.8 & -0.4 \\ -12 & 5 & -4.9 & 5.3 & -1.7 & 0.3 \\ 3.7 & -1.5 & 1.8 & -1.7 & 0.5 & -0.1 \\ -0.7 & 0.2 & -0.4 & 0.3 & -0.1 & 0.02 \end{bmatrix}, K_3 = \begin{bmatrix} -0.4 \\ -0.02 \\ -0.2 \\ -2.7 \\ -8.9 \\ -21.4 \end{bmatrix},$$

$\gamma_3 = 11.4;$

$$X_4 = \begin{bmatrix} 16.9 & -7.6 & 7.9 & -7.2 & 2.1 & -0.3 \\ -7.6 & 4.6 & -3 & 3.1 & -0.9 & 0.1 \\ 7.9 & -3 & 11.4 & -3.7 & 1.3 & -0.3 \\ -7.2 & 3.1 & -3.7 & 3.1 & -0.9 & 0.1 \\ 2.1 & -0.9 & 1.3 & -0.9 & 0.3 & -0.05 \\ -0.3 & 0.1 & -0.3 & 0.1 & -0.05 & 0.01 \end{bmatrix}, K_4 = \begin{bmatrix} -0.4 \\ -0.02 \\ -0.2 \\ -3.3 \\ -11.5 \\ -29.5 \end{bmatrix},$$

$\gamma_4 = 10.5.$

If we consider $0 \leq u \leq 20$, the following Lyapunov matrices and feedback gains are (controller 3):

$\mu = 0.18;$

$$X_1 = \begin{bmatrix} 43.6 & -21 & 12.1 & -18.2 & 5.2 & -0.9 \\ -21 & 14.2 & -4.9 & 8.4 & -2.3 & 0.3 \\ 12.1 & -4.9 & 10.9 & -5.6 & 1.9 & -0.4 \\ -18.2 & 8.4 & -5.6 & 7.8 & -2.2 & 0.4 \\ 5.2 & -2.3 & 1.9 & -2.2 & 0.6 & -0.1 \\ -0.9 & 0.3 & -0.4 & 0.4 & -0.1 & 0.02 \end{bmatrix}, K_1 = \begin{bmatrix} -0.6 \\ -0.02 \\ -0.4 \\ -4.1 \\ -15.4 \\ -42.3 \end{bmatrix},$$

$\gamma_1 = 9.6;$

$$X_2 = \begin{bmatrix} 26.6 & -13. & 9.1 & -10.8 & 2.9 & -0.4 \\ -13.5 & 10.5 & -3.8 & 5.2 & -1.3 & 0.2 \\ 9.1 & -3.8 & 10.4 & -4.1 & 1.3 & -0.3 \\ -10.8 & 5.2 & -4.1 & 4.5 & -1.2 & 0.2 \\ 2.9 & -1.3 & 1.3 & -1.2 & 0.3 & -0.06 \\ -0.4 & 0.2 & -0.3 & 0.2 & -0.06 & 0.01 \end{bmatrix},$$

$$K_2 = \begin{bmatrix} -0.7 \\ -0.01 \\ -0.4 \\ -4.9 \\ -19.8 \\ -57.9 \end{bmatrix}, \gamma_2 = 8.9;$$

$$X_3 = \begin{bmatrix} 43.4 & -20.9 & 12.1 & -18.1 & 5.1 & -0.8 \\ -20.9 & 14.1 & -4.8 & 8.4 & -2.2 & 0.3 \\ 12.1 & -4.8 & 10.8 & -5.5 & 1.9 & -0.4 \\ -18.1 & 8.4 & -5.5 & 7.7 & -2.2 & 0.4 \\ 5.1 & -2.2 & 1.9 & -2.2 & 0.6 & -0.1 \\ -0.8 & 0.3 & -0.4 & 0.4 & -0.1 & 0.02 \end{bmatrix},$$

$$K_3 = \begin{bmatrix} -0.6 \\ -0.02 \\ -0.4 \\ -4 \\ -15 \\ -42 \end{bmatrix}, \gamma_3 = 9.6;$$

$$X_4 = \begin{bmatrix} 26.5 & -13.5 & 9 & -10.8 & 2.9 & -0.4 \\ -13.5 & 10.5 & -3.8 & 5.2 & -1.3 & 0.2 \\ 9 & -3.8 & 10.4 & -4.1 & 1.3 & -0.3 \\ -10.8 & 5.2 & -4.1 & 4.5 & -1.2 & 0.2 \\ 2.9 & -1.3 & 1.3 & -1.2 & 0.3 & -0.06 \\ -0.4 & 0.2 & -0.3 & 0.2 & -0.06 & 0.01 \end{bmatrix},$$

$$K_4 = \begin{bmatrix} -0.7 \\ -0.02 \\ -0.4 \\ -4.8 \\ -19.7 \\ -57.7 \end{bmatrix}, \gamma_4 = 8.9.$$

The desired glucose concentration level is considered $G_{op} = 80\ mg/dL$ and to model the *10, 20* and *30 gr* meal glucose disturbances, the exponential function $v = \alpha\ e^{-0.05t}$ with $\alpha = 1,2,3$ is exerted to the system at $t = 0$ min (Fig.1).

The Bergman model closed-loop output and control signal considering the various range of input saturation constraints for $\alpha = 1,2,3$ are represented in Figs. 2-7. As shown in Figs. 2, 4 and 6 the fuzzy $H_\infty$ controllers can significantly control the blood glucose concentration level, subjected to meal disturbance, while the input saturation limitation is satisfied. Furthermore, the disturbance amplitude influences the plasma sugar level's peak and is also influenced by the input saturation constraint directly. It means that the less the input saturation constraint is, the better and faster the controller performs and the better $H_\infty$ performance is obtained.

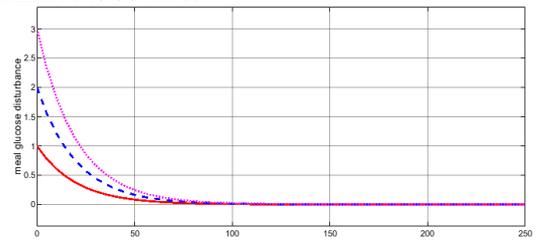

Fig. 1. The consumed glucose ($\alpha = 1$ by solid line $\alpha = 2$ by dashed line, $\alpha = 3$ by dotted line)

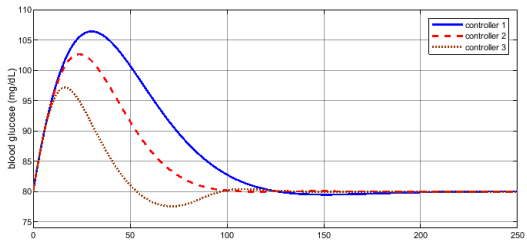

Fig. 2. The minimal model plasma glucose level for α =1(controller 1 by solid line, controller 2 by dashed line, controller 3 by dotted line)

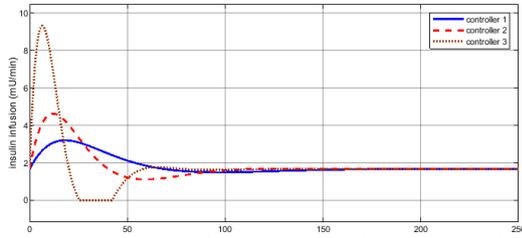

Fig. 3. The minimal model rate of insulin injection for α =1 (controller 1 by solid line, controller 2 by dashed line, controller 3 by dotted line)

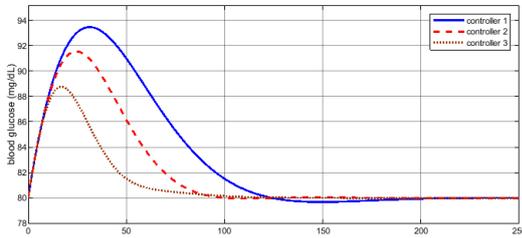

Fig. 4. The minimal model plasma glucose level for α =2 (controller 1 by solid line, controller 2 by dashed line, controller 3 by dotted line)

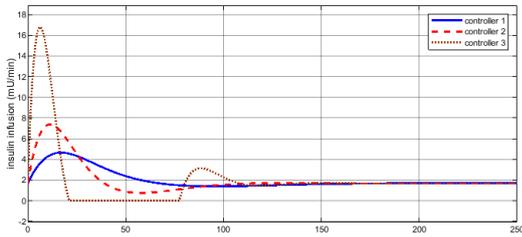

Fig. 5. The minimal model rate of insulin injection for α =2 (controller 1 by solid line, controller 2 by dashed line, controller 3 by dotted line)

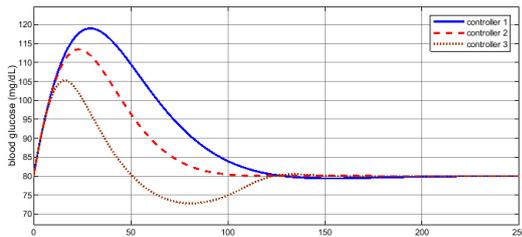

Fig. 6. The minimal model plasma glucose level for α =3 (controller 1 by solid line, controller 2 by dashed line, controller 3 by dotted line)

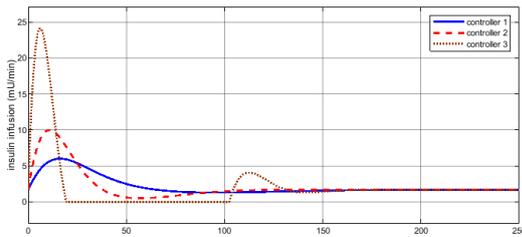

Fig. 7. The minimal model rate of insulin injection for α =3 (controller 1 by solid line, controller 2 by dashed line, controller 3 by dotted line)

As can be seen in Figs. 2-7 the cost of better control performance (faster response and lower blood glucose concentration peak) is the less input saturation limit and the more insulin infusion rate. The applicability of the suggested $H_\infty$ fuzzy model-based controller is that for the different regarded meal glucose disturbances, the insulin infusion rates do not reach the upper input limit; therefore, the suggested controller has better $H_\infty$ performance practically.

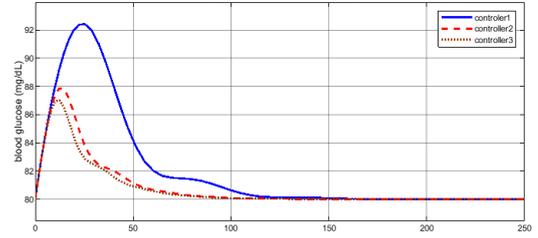

Fig. 8. The Tolic model plasma glucose level for α =1(controller 1 by solid line, controller 2 by dashed line, controller 3 by dotted line)

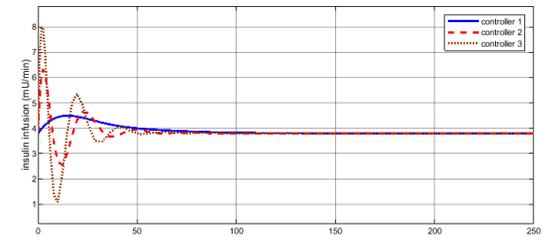

Fig. 9. The Tolic model rate of insulin injection for α =1 (controller 1 by solid line, controller 2 by dashed line, controller 3 by dotted line)

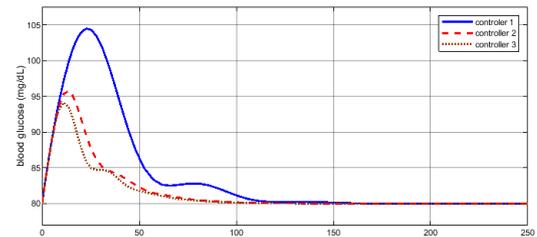

Fig. 10. The Tolic model plasma glucose level for α =2 (controller 1 by solid line, controller 2 by dashed line, controller 3 by dotted line)

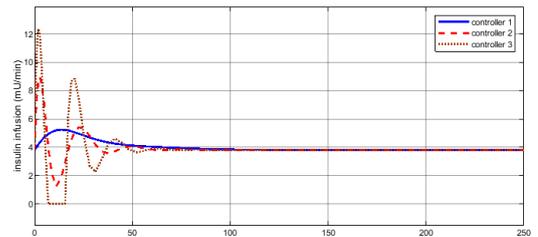

Fig. 11. The Tolic model rate of insulin injection for α =2 (controller 1 by solid line, controller 2 by dashed line, controller 3 by dotted line)

Figs. 8, 10, and 12 indicate the closed-loop output, and Figs. 9, 11, and 13 show the control effort in the presence of disturbance *(α =1,2,3)* in the Tolic model. As it is shown, the suggested controller can successfully decline the disturbance effect on output and converge blood glucose level to its basal value. Compared to the Bergman system, the closed-loop Tolic system is faster with smaller variations with the desired value of blood glucose concentration, additionally, control efforts in Figs. 9, 11, and 13 demonstrate more oscillatory behavior more similar to pancreas behavior.

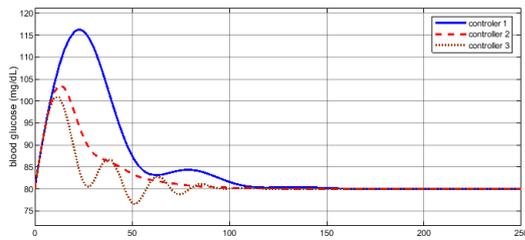

Fig. 12 The Tolic model plasma glucose level for *α* =3 (controller 1 by *solid* line, controller 2 by dashed line, controller 3 by dotted line)

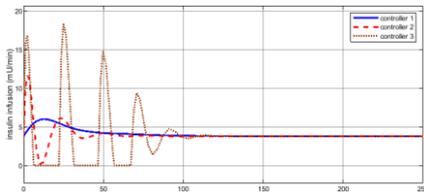

Fig. 13. The Tolic model rate of insulin injection for *α* =3 (controller 1 by solid line, controller 2 by dashed line, controller 3 by dotted line)

## IV. Conclusion

We have proposed an $H_\infty$ Takagi-Sugeno (TS) fuzzy model-based controller to minimize the consumed sugar disturbance effect and keeping blood glucose level on its standard level. In the controller design procedure, the insulin infusion pump limitation is considered to design a more practical controller and develop the proposed controller to different pumps' power.

Firstly, based on the SNA, the TS fuzzy approximation systems of the nonlinear models of glucose insulin system were obtained. The TS fuzzy models contain the sum of some weighted linear subsystems that enable us to use linear control theory in the nonlinear systems. Secondly, the LMI-based controller design theories were employed to achieve $H_\infty$ performance criteria and closed-loop stability, while different insulin infusion pump limitations can be considered. Finally, State feedback controllers for each linear subsystem were calculated via convex numerical techniques. Simulation results verify that the designed controllers, within different control signal limits, effectively mitigate disturbances in the original Bergman and Tolic systems.